\def\year{2022}\relax
\documentclass[letterpaper]{article} 
\usepackage{aaai22}  
\usepackage{times}  
\usepackage{helvet}  
\usepackage{courier}  
\usepackage[hyphens]{url}  
\usepackage{graphicx} 
\usepackage{natbib}  
\usepackage{caption} 
\DeclareCaptionStyle{ruled}{labelfont=normalfont,labelsep=colon,strut=off} 
\usepackage{algorithm}
\usepackage{algorithmic}
\usepackage{amsmath}
\usepackage{amsfonts}
\usepackage{bibentry}
\usepackage{newfloat}
\usepackage{listings}
\floatstyle{ruled}
\newfloat{listing}{tb}{lst}{}
\floatname{listing}{Listing}

\title{Representation Learning with Graph Neural Networks for \\ Speech Emotion Recognition}
\author{
    Junghun Kim and Jihie Kim
}
\affiliations{
Department of Artificial Intelligence, Dongguk University, Seoul, Korea \\ {riseup32@dongguk.edu, jihie.kim@dgu.edu}

}

\begin{document}

\maketitle

\begin{abstract}
Learning expressive representation is crucial in deep learning. In speech emotion recognition (SER), vacuum regions or noises in the speech interfere with expressive representation learning. 
However, traditional RNN-based models are susceptible to such noise. 
Recently, Graph Neural Network (GNN) has demonstrated its effectiveness for representation learning, and we adopt this framework for SER. 
In particular, we propose a cosine similarity-based graph as an ideal graph structure for representation learning in SER. 
We present a Cosine similarity-based Graph Convolutional Network (CoGCN) that is robust to perturbation and noise. 
Experimental results show that our method outperforms state-of-the-art methods or provides competitive results with a significant model size reduction with only 1/30 parameters.
\end{abstract}

\section{Introduction}

The main challenge in Speech Emotion Recognition (SER) is to learn utterance-level representations from frame-level features, which starts with learning expressive frame-level representations.
Previous works \citep{zadeh2017tensor, zadeh2018memory, zhao2019attention} mainly used LSTM \citep{hochreiter1997long} or GRU \citep{chung2014empirical}, a variant of RNN, to handle this challenge.
However, RNN-based models can only consider left-to-right or right-to-left information. 
Even with bidirectional RNN that is learned for each direction and concatenated, both directions' information cannot be considered simultaneously. 
RNN architecture also propagates irrelevant information continuously, even if there is noise.
Therefore, surrounding frame information cannot be learned properly, and frame-level representations are perturbated easily.
To overcome these issues, we propose using a GNN architecture, as it can relate information in various parts simultaneously.
Furthermore, GNN can also optimize the model size due to parameter sharing, which extends SER’s applications that require limited memory space, such as on-device speech recognition.
There are many multimodal \citep{aguilar2019multimodal, priyasad2020attention} or multi-task \citep{li2019improved, latif2020multi} approaches in SER, but we focus on unimodal and single task for in this study to demonstrate the effectiveness of GNN for SER.

The utterance can be represented as a noisy graph structure in which voice frame and vacuum (with no voice) frame coexist. 
Therefore, it is important to ensure that useful (with-voice) frames are not perturbed by irrelevant (no-voice) frames.
An ideal graph structure can help with it.
If the input graphs are ideal, irrelevant information can be filtered out through the message passing process. 
To construct an ideal graph structure, we propose to use a cosine similarity metric.
The superiority of the cosine similarity metric in graph construction has already been demonstrated in other applications \citep{chen2020iterative}.
We adopt it in SER as we believe that graph structures constructed through the cosine similarity metric are more robust to perturbation by pruning out neighbors that have irrelevant information.

Furthermore, we design our GNN architecture using a message passing framework of the Graph Convolutional Network (GCN) \citep{kipf2016semi} with the cosine similarity-based graph structure.
We also construct additional modules in our architecture, such as an acoustic pre-processing layer and a skip connection.
The proposed GNN architecture better captures neighbors' information and enables expressive frame-level representation learning.

\begin{figure*}[th]
\centering
    \includegraphics[width=1\linewidth]{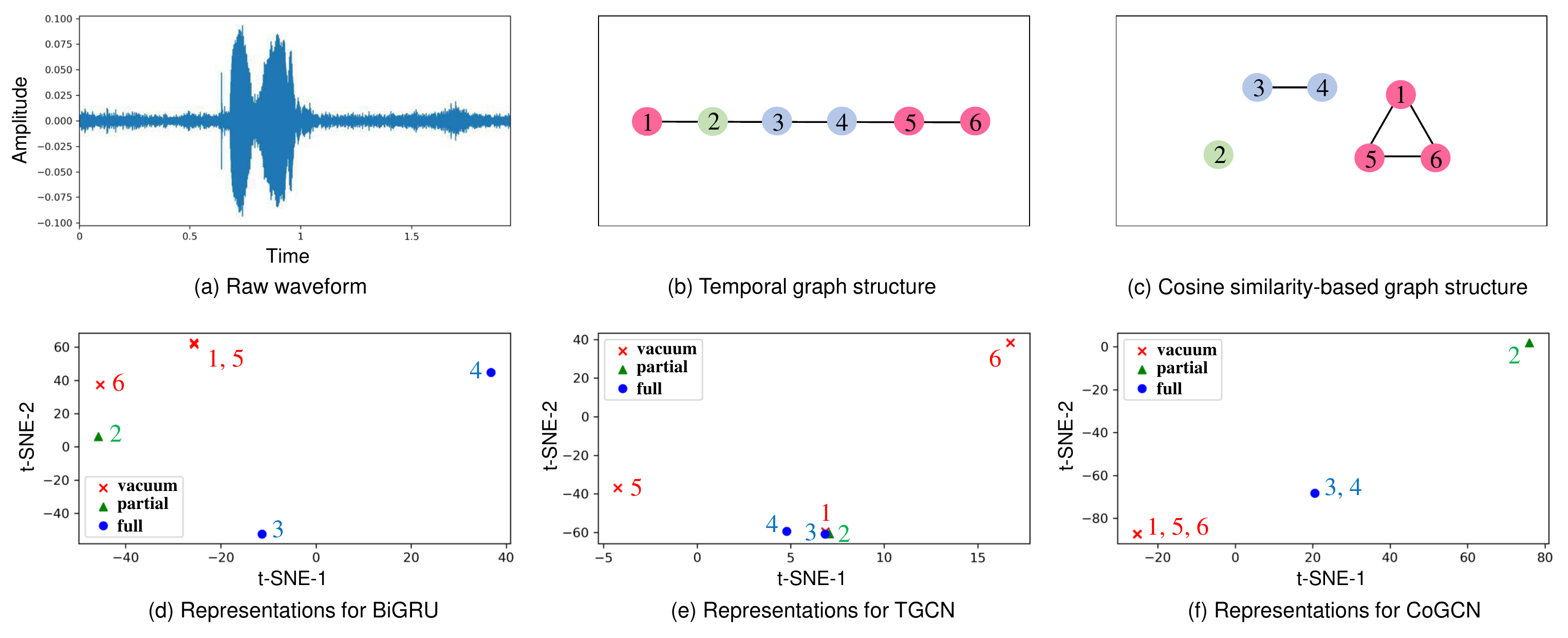}
\caption{Top: Temporal graph structure and cosine similarity-based graph structure for the utterance sample. 
Bottom: t-SNE \citep{van2008visualizing} visualization of the representations for each method.}
\end{figure*}

Figure 1 gives an overview of SER representations with graphs.
Figures 1 (b) and 1 (c) show a temporal graph structure and a cosine similarity-based graph structure for the utterance sample in Figure 1 (a). 
The red node is a vacuum (no-voice) node, the green node is a partial-voice (a part of the frame is voice) node, and the blue node is a full-voice (the entire frame contains voice) node, using 50ms frame window size and 25ms frame intervals.
We compare the representations learned with Bidirectional GRU (BiGRU) and the GCN's representations learned from each graph structure. 
In Figure 1 (d), nodes 3 and 4, which should be used in prediction, are far from each other.
This means that the representations for the prediction are not properly learned.
In Figure 1 (e), nodes 3 and 4 are mapped to a nearby latent space, but nodes 1, 2, and 3 are also mapped closely due to the perturbation. 
In Figure 1 (f), the nodes with similar characteristics are mapped closely in the latent space, which means that our proposed method can be robust to perturbation.

Our contributions are summarized as follows: 
\begin{itemize}
\item We propose a cosine similarity-based graph structure as an ideal graph structure for SER.
\item We present a Cosine similarity-based Graph Convolutional Network (CoGCN), as a GCN variant for SER.
\item Experimental results show that our method outperforms state-of-the-art methods or provides competitive results with a significant model size reduction with only 1/30 parameters.
\end{itemize}

\section{Related Work}

GNN learns node representations from neighbors through message passing and aggregation. 
\citep{kipf2016semi} proposed Graph Convolutional Network (GCN), inspired by the first-order graph Laplacian methods.
\citep{hamilton2017inductive} proposed GraphSAGE (SAmple and aggreGatE) sampling a fixed number of neighbors to keep the computational complexity consistent.
\citep{velivckovic2017graph} proposed Graph Attention Network (GAT) to allocate different weights to neighbors.
\citep{xu2018powerful} developed Graph Isomorphism Network (GIN) that is probably the most expressive among GNN varients.
Generally, GIN's message passing method learns expressive representation, but it does not work well for SER because the sum aggregation over the multiset including noises can disturb the representation learning.
We design our GNN architecture using GCN's message passing method since we believe that it aggregates abundant distribution information in SER that values statistical information.
Furthermore, we construct additional modules along the design space guidelines for well-performing GNN of \citep{you2020design}.

To learn the utterance-level representations from the frame-level features, \citep{latif2019direct, peng2020speech} used the RNN based-model.
CLDNN \citep{latif2019direct} used a combination of CNN and LSTM to complement each architecture's shortcomings.
ASRNN \citep{peng2020speech} used the BiLSTM and attention mechanism to strengthen the time step importance.
Recently, CA-GRU \citep{su2020improving} introduced the GNN architecture in SER, where frame-level representations encoded in BiGRU are used as node features of GNN architecture.
However, frame-level representations trained in this architecture can be easily perturbed and require many parameters.
To make SER more robust to perturbation, we use a Fully Connected (FC) layer as a pre-processing layer instead of GRU.
We also use a cosine similarity metric that does not require many additional parameters.

\section{Approach}
We first describe notations used for describing our method.
Then we present the graph structure constructed through the cosine similarity metric.
Finally, we present Cosine similarity-based Graph Convolutional Network (CoGCN), a GCN variant. 

\subsection{Notations}
We begin by summarizing the notations used in the GNN architecture.
Let $G=(V, E, X)$ denote a graph, where $V$ is a vertex set, $E$ is an edge set, $X \in \mathbb{R}^{n \times d}$ is a feature matrix, and $n$ and $d$ are the number of vertices and the dimension of the feature vector, respectively.
$A \in \mathbb{R}^{n \times n}$ is an adjacency matrix and $\mathcal{N}(i)$ is a set of neighbors of node $i$.
Therefore, given a set of graphs $\{G_1, ..., G_N\}$ and their labels $\{y_1, ..., y_N\}$, we aim to learn a graph representation vector $h_G$ to predict the label of the entire graph.

\subsection{Graph Structure}
Since the utterance is sequential data, it has a basic temporal graph structure where the center node has edges on both sides.
The temporal graph structure is not ideal because it is difficult to capture long-term dependencies and is easily perturbed by irrelevant neighbors.
In this paper, we propose a graph structure constructed with the cosine similarity metric as an ideal graph structure for SER.
Cosine similarity-based graph structure can capture long-term dependencies and prevent perturbation from irrelevant neighbors. 
The process of generating the cosine similarity-based graph is as follows:
$$
s_{ij}=\frac{x_i^Tx_j}{||x_i||\times||x_j||},
\eqno{(1)}
$$

$$
a_{ij}=\begin{cases}1, & \mathrm{if} \; s_{ij} \geq \gamma
\\ 
0, & \mathrm{otherwise,} \end{cases}
\eqno{(2)}
$$
where $x_i \in \mathbb{R}^{d}$ is the $i$-th row of the feature matrix $X$, $\gamma$ is threshold hyperparameter, $s_{ij}$ is the cosine similarity between node $i$ and node $j$, and $a_{ij}$ is the corresponding element of the adjacency matrix $A$.

\subsection{Model Architecture}

\begin{figure}[t]
\centering
    \includegraphics[width=1\linewidth]{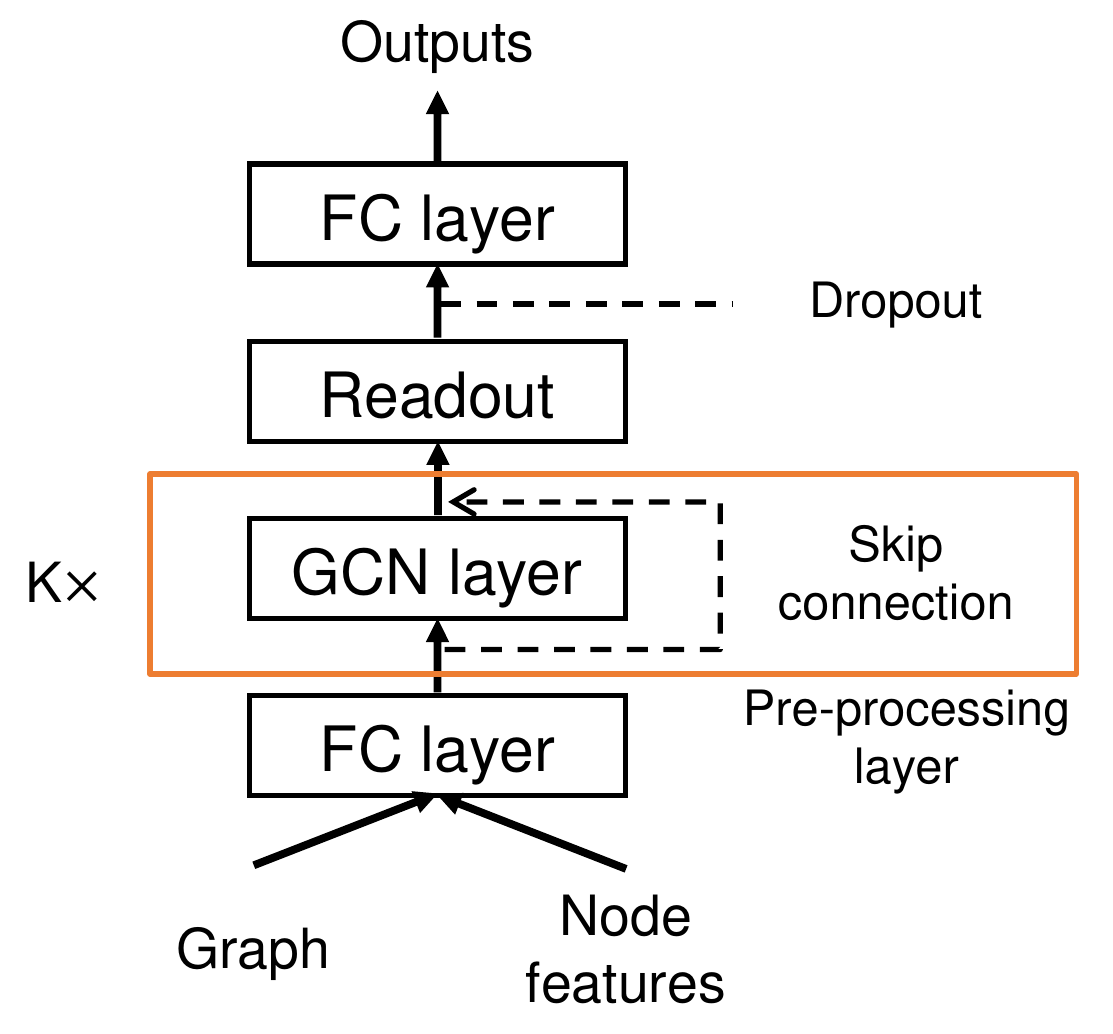}
\caption{Model architecture.}
\end{figure}

There are numerous GNN variants based on the message passing method used.
GIN's message passing method learns expressive representation in many tasks but does not work well for SER because the sum aggregation over the multiset including noises can disturb the representation learning.
Instead, we select the GCN's message passing method that best matches our task since we believe that it aggregates abundant distribution information in SER that values statistical information.
\citep{you2020design} provided comprehensive guidelines about design spaces \{Batch Normalization, Dropout, Activation, Aggregation, Layer connectivity, Pre-processing Layers, Message passing Layers, Post-processing Layers\} for designing a well-performing GNN.
We performed experiments with possible design spaces along the guidelines and found that one FC pre-processing layer and skip connection helped the most in improving the performance.
Therefore, the node representation is calculated as follows:

$$
h_i^{(0)}=\mathrm{ReLU}(W_px_i+b_p),
\eqno{(3)}
$$

$$
h_i^{\prime^{(k+1)}}=\mathrm{ReLU}(W_e^{(k)}\sum_{j}\frac{1}{\sqrt{\hat{d_i}\hat{d_j}}}h_j^{(k)}),
\eqno{(4)}
$$

$$
h_i^{(k+1)}=h_i^{\prime^{(k+1)}}+h_i^{(k)},
\eqno{(5)}
$$
where $\hat{d_i}=1+\sum_{j \in \mathcal{N}(i)}a_{ij}$, $W_p \in \mathbb{R}^{z \times d}$ and $W_e^{(k)} \in \mathbb{R}^{z \times z}$ are the learnable weight matrices, and $b_p \in \mathbb{\mathbb{R}}^{z}$ is the learnable bias vector.
$h_i^{(k)}$ is the node representation of the $k$-th layer of node $i$ for $k=1,...,K$, and $z$ is the number of hidden units.

Node representation cannot be used directly for the graph classification task. 
Therefore, given the final iteration node representations, we use the readout function to produce a graph representation.
Finally, the graph label is predicted through the FC layer followed by softmax activation that takes the graph representation as the input.

$$
h_G=\mathrm{READOUT}({h_v^{(K)}\mid v \in G}),
\eqno{(6)}
$$

$$
\hat{y}_G=\mathrm{softmax}(W_oh_G+b_o),
\eqno{(7)}
$$
where $W_o \in \mathbb{R}^{C \times z}$ is the learnable weight matrix, $b_o \in \mathbb{R}^{C}$ is the learnable bias vector, and $C$ is the number of classes.

$\mathrm{READOUT}$ can be a simple permutation-invariant function, such as summation or mean, or max-pooling. 
We believe that the graph's statistical and distributional information is important in this study, so we use the mean-pooling as the readout function, following \citep{xu2018powerful}'s suggestion.
The overall architecture is illustrated in Figure 2.

\section{Experiments}

\subsection{Acoustic Features}
Frame-level features are extracted from raw waveforms using the openSMILE \citep{eyben2010opensmile} speech toolkit with 25ms frame window size and 10ms frame intervals. 
We use the extended Geneva Minimalistic Acoustic Parameter Set (eGeMAPS) introduced by \citep{eyben2015geneva} to extract frame-level features with a total 88-dimension.

\subsection{Dataset}
An IEMOCAP \citep{busso2008iemocap} consists of 5 sessions, and each session includes 2 actors (1 male and 1 female). 
It consists of 9 emotions, but in this paper, following the previous studies, data with only 4 emotions \{neutrality, happiness (including excited), sadness, and anger\} is used, which is 5531 utterances.

\subsection{Experiments Setup}

\begin{table}[t]
\centering
\begin{tabular}{cccc}
\textbf{Method} & \textbf{\# parameter} & \textbf{WA} & \textbf{UA}\\
\hline
BiGRU & 464K & 59.76 & 59.71 \\
\hline
TGCN & 56K & 61.52 & 62.43 \\
w/o skip & 56K & 60.97 & 61.77 \\
w/o skip, pre & 45K & 60.61 & 61.50 \\
\hline
CoGCN & 56K & \textbf{62.64} & \textbf{63.67} \\
w/o skip & 56K & 61.35 & 62.56 \\
w/o skip, pre & 45K & 61.14 & 62.34 \\
\hline
\end{tabular}
\caption{\label{citation-guide}
Comparison of different combinations.
}
\end{table}

Each method has $K \in$ \{2, 3, 4\} layers with 128 hidden units, and dropout is applied with p = 0.1 after the readout function.
$K$ is selected through the validation set results.
We train all the models for a maximum of 50 epochs with a batch size of 32 using the Adam optimizer \citep{kingma2014adam} with a learning rate of 1e-3. 
We search the threshold parameter $\gamma$ in \{0.5, 0.55, 0.6\}. 
Finally, following the state-of-the-art methods' experimental settings being compared, we perform leave-one-person-out cross-validation and report the average of Weighted Accuracy (WA) and Unweighted Accuracy (UA).

\subsection{Result Analysis}

\begin{figure}[t]
\centering
    \includegraphics[width=1\linewidth]{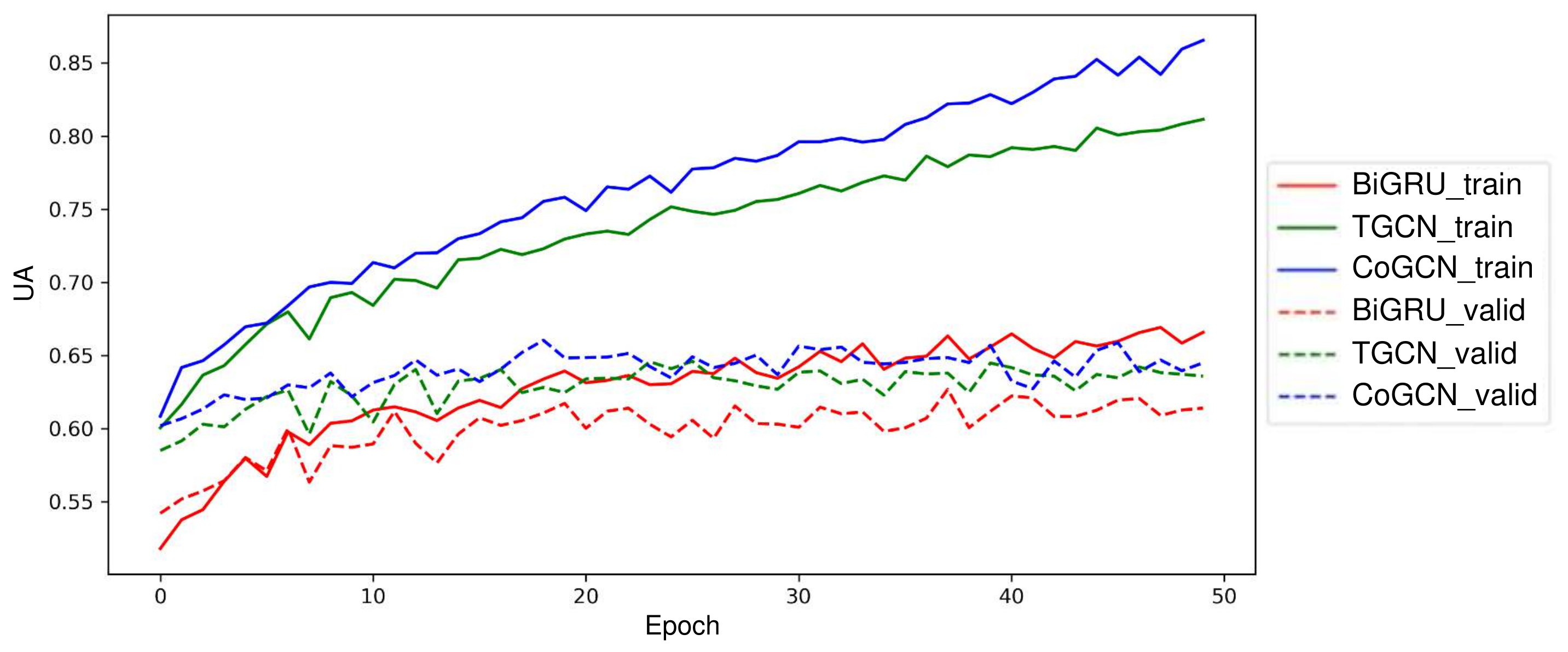}
\caption{Training set and validation set learning curve of BiGRU, TGCN, and CoGCN.}
\end{figure}

\begin{table}[t]
\small
\centering
\begin{tabular}{cccc}
\textbf{Method} & \textbf{\# parameter} & \textbf{WA} & \textbf{UA}\\
\hline
CLDNN \citep{latif2019direct} & 250K & - & 60.23 \\
ASRNN \citep{peng2020speech} & 6M & - & 62.60 \\
CA-GRU \citep{su2020improving} & 1.6M & 62.27 & \textbf{63.80} \\
\hline
CoGCN (Ours) & \textbf{56K} & \textbf{62.64} & 63.67 \\
\hline
\end{tabular}
\caption{\label{citation-guide}
Summary of results in terms of WA and UA.
}
\end{table}

To demonstrate the effect of our proposed approach, we compare CoGCN with BiGRU and Temporal graph based GCN (TGCN) learned under the same conditions.
Table 1 shows the study results with BiGRU, TGCN, and CoGCN.
We can see the performance improvement with GNN over RNN (BiGRU).
We can also see the benefit of cosine similarity-based graph structure when compared with temporal graph structure.
When skip connection is removed from the GNN architectures, the performance drops in general.
Besides, when the pre-processing layer is removed from GNN architectures without skip connection, we see an additional performance drop.

In Figure 3, each method's learning curve with the training set and the validation set supports the findings above.
Intuitively, BiGRU is relatively underfitting, and CoGCN learns more expressive node representations than TGCN.

Table 2 shows a comparison with the state-of-the-art methods.
The dash symbol denotes that reported results do not exist.
Our method outperforms CLDNN, ASRNN and achieves a competitive performance when compared to CA-GRU.
It is noteworthy that our approach significantly reduces the model size and the number of parameters is only 1/30 of CA-GRU.
Such results can help applications that require limited memory space, such as on-device speech recognition.

\section{Conclusions}
In this paper, we propose a cosine similarity-based graph structure as an ideal graph structure for SER, and present the CoGCN, as a GCN variant for SER.
Finally, we show that our method outperforms state-of-the-art methods or provides competitive results with a significant model size reduction with only 1/30 parameters.

\section{Acknowledgement}
This research was supported by the MSIT(Ministry of Science and ICT), Korea, under the ITRC(Information Technology Research Center) support program(IITP-2021-2020-0-01789) supervised by the IITP(Institute for Information \& Communications Technology Planning \& Evaluation).

\bibliography{aaai22}

\end{document}